\documentclass[12pt,a4paper]{article}
\usepackage{graphicx}
\usepackage{amsmath}
\usepackage{amssymb}
\usepackage{epsfig}
\usepackage{times}
\begin{document}

\title{Einstein's Contributions to Quantum Theory\footnote{Invited talk at \textit{Hadron
Collider Physics Symposium} (HCP2005) July 4-9 2005, Les Diablerets, Switzerland}}
\author{
Norbert Straumann\\
Institute for Theoretical Physics\\
University of Z\"urich, Switzerland}

\maketitle

\begin{abstract}
\noindent Einstein's revolutionary light quantum hypothesis of 1905 and his further
contributions to quantum theory are reviewed.
\end{abstract}

\section{Introduction}
During this \emph{World Year of Physics} physicists celebrate all over the world the
astounding sequence of papers that Einstein wrote in rapid succession during the year 1905.
But already before this \emph{annus mirabilis} Einstein had published remarkable papers in
the \textit{Annalen der Physik}, the journal to which he submitted most of his early work.
Of crucial importance for his further research were three papers on the foundations of
statistical mechanics, in which he tried to fill what he considered to be a gap in the
mechanical foundations of thermodynamics. At the time when Einstein wrote his three papers
he was not familiar with the work of Gibbs and only partially with that of Boltzmann.
Einstein's papers form a bridge, parallel to the \textit{Elementary Principles of
Statistical Mechanics} by Gibbs in 1902, between Boltzmann's work and the modern approach
to statistical mechanics. In particular, Einstein independently formulated the distinction
between the microcanonical and canonical ensembles and derived the equilibrium distribution
for the canonical ensemble from the microcanonical distribution. Of special importance for
his later research was the derivation of the energy-fluctuation formula for the canonical
ensemble.

Einstein's profound insight into the nature and size of fluctuations played a decisive role
for his most revolutionary contribution to physics: the light-quantum hypothesis. Indeed,
Einstein extracted the light-quantum postulate from a statistical-mechanical analogy
between radiation in the Wien regime\footnote{The `Wien regime' corresponds to high
frequency and/or low temperature, such that $h\nu\gg kT$, where $h$ and $k$ are Planck's
and Boltzmann's constants respectively.} and a classical ideal gas of material particles.
In this consideration Boltzmann's principle, relating entropy and probability of
macroscopic states, played a key role. Later Einstein extended these considerations to an
analysis of energy and momentum fluctuations of the radiation field. For the latter he was
also drawing on ideas and methods he had developed in the course of his work on Brownian
motion, another beautiful application of fluctuation theory. This definitely established
the reality of atoms and molecules, and, more generally, gave strong support for the
molecular-kinetic theory of thermodynamics.

Fluctuations also played a prominent role in Einstein's beautiful work on critical
opalescence. Many years later he applied this magic wand once more to gases of identical
particles, satisfying the Bose-Einstein statistics. With this work in 1924 he extended the
particle-wave duality for photons to massive particles. It is well-known that
Schr\"{o}dinger was much stimulated by this profound insight. As an application, Einstein
also discovered what is known as Bose-Einstein condensation, that has become a very topical
research field.

\section{Einstein's first paper from 1905}
The generations of physicists that learned quantum theory after the great breakthrough in
1925-26 rarely know about the pioneering role of Einstein in the development of this field
during the previous twenty years. With his work on quantum theory alone he would already
belong to the central figures of twentieth century physics. In the first of his 1905 papers
he introduced the hypothesis of light quanta, a step that he considered himself as his only
revolutionary one. The course of physics would presumably have been quite different without
this rather bold suggestion. Indeed, Einstein was the first who clearly realized that the
empirical energy distribution of the black-body radiation was in dramatic conflict with
classical physics, and thus a radically different conception of radiation was required.
Most physicists reduce the content of Einstein's paper ``On a heuristic point of view
concerning the production and transformation of light'' to what he wrote about the
photoelectric effect. This was, however, just an important application of a much more
profound analysis, that he soon supplemented in various ways.

We begin by briefly reviewing the line of thought of the March paper (CPAE Vol.\,2,
Doc.\,14) ``whose significance and originality can hardly be overestimated'' (Res Jost).
In a first section Einstein emphasizes that classical physics inevitably leads to a
nonsensical energy distribution for black-body radiation, but that the spectral
distribution, $\rho(T,\nu)$, must approximately be correct for large wavelengths and
radiation densities (classical regime).\footnote{This is, to our knowledge, the first
proposal of a `correspondence argument', which is of great heuristic power, as we will
see.} Applying the equipartition theorem for a system of resonators (harmonic oscillators)
in thermal equilibrium, he found independently what is now known as the Rayleigh-Jeans
law\footnote{Einstein uses the following relation between $\rho(T,\nu)$ and the mean
oscillator energy $\bar{E}(T,\nu)$ at temperatur $T$, found by Planck:
$\rho(T,\nu)=\frac{8\pi\nu^2}{c^3}\bar{E}(T,\nu)$.}: $\rho(\nu,T)=(8\pi\nu^2/c^3)kT$.
Einstein stresses that this law ``not only fails to agree with experience (...), but is out
of question'' because it implies a diverging total energy density (ultraviolet
catastrophe). In a second section he then states that the Planck formula ``which has been
sufficient to account for all observations made so far'' agrees with the classically
derived formula in the mentioned limiting domain for the following value of the Avogadro
number
\begin{equation}
N_A=6.17\times 10^{23}\,.
\end{equation}
This value was already found by Planck, though not using a correspondence argument, but
rather relying on the strict validity of his formula and the assumptions that led to its
derivation. Einstein's correspondence argument now showed ``that Planck's determination of
the elementary quanta is to some extent independent of his theory of black-body
radiation.'' Indeed, Einstein understood from first principles exactly what he did. A
similar correspondence argument was used by him more than ten years later in his famous
derivation of Planck's formula (more about this later). Einstein concludes these
considerations with the following words:
\begin{quote}
``\textit{The greater the energy density and the wavelength of the radiation, the more
useful the theoretical principles we have been using prove to be; however, these principles
fail completely in the case of small wavelengths and small radiation densities.''}
\end{quote}

Einstein now begins to analyze what can be learned about the structure of radiation from
the empirical behavior in the Wien regime, i.e., from Wien's radiation formula for the
spectral energy-density
\begin{equation}
\label{eq:LQH-Wien} \rho(T,\nu)=\frac{8\pi\nu^2}{c^3}h\nu e^{-h\nu/kT}\,.
\end{equation}
Let $E_V(T,\nu)$ be the energy of radiation contained in the volume $V$ and within the
frequency interval $[\nu\,,\,\nu+\Delta\nu]$ ($\Delta\nu$ small), that is,
\begin{equation}
\label{eq:LQH-Energy} E_V(T,\nu)=\rho(T,\nu)\,V\,\Delta\nu\,.
\end{equation}
and, correspondingly, $S_V(T,\nu)=\sigma(T,\nu)\,V\,\Delta\nu$ for the entropy.
Thermodynamics now implies
\begin{equation}
\label{eq:LQH-Step1} \frac{\partial\sigma}{\partial\rho}=\frac{1}{T}\,.
\end{equation}
Solving (\ref{eq:LQH-Wien}) for $1/T$ and inserting this into (\ref{eq:LQH-Step1}) gives
\begin{equation}
\label{eq:LQH-Step2} \frac{\partial\sigma}{\partial\rho}
=-\frac{k}{h\nu}\ln\left[\frac{\rho}{8\pi h\nu^3/c^3}\right]\,.
\end{equation}
Integration yields
\begin{equation}
\label{eq:LQH-Step3} S_V=-k\frac{E_V}{h\nu}\left\{ \ln\left[\frac{E_V}{V\Delta\nu\,8\pi
h\nu^3/c^3}\right]-1\right\}\,.
\end{equation}
In his first paper on this subject, Einstein focused attention to the volume dependence of
radiation entropy, as displayed by this expression. Fixing the amount of energy, $E=E_V,$
one obtains
\begin{equation}
\label{eq:LQH-Step4} S_V-S_{V_0}=k\frac{E}{h\nu}\ln\left(\frac{V}{V_0}\right)=
k\ln\left(\frac{V}{V_0}\right)^{E/h\nu}\,.
\end{equation}

So far only thermodynamics has been used. Now Einstein brings into the game what he called
Boltzmann's principle, which was already of central importance in his papers on statistical
mechanics. According to Boltzmann, the entropy $S$ of a system is connected  with the
number of possibilities $W$, by which a macroscopic state can microscopically be realized,
through the relation
\begin{equation}
\label{eq:LQH-BoltzmanPrinzip} S=k\ln W\,.
\end{equation}
In a separate section Einstein recalls this fundamental relation between entropy and
``statistical probability'' (Einstein's terminology), before applying it to an ideal gas of
$N$ particles in volumes $V$ and $V_0$, respectively. For the relative probability of the
two situations one has
\begin{equation}
\label{eq:LQH-BoltzmanAppl} W=\left(\frac{V}{V_0}\right)^N\,,
\end{equation}
and hence for the entropies
\begin{equation}
\label{eq:LQH-Step5} S(V,T)-S(V_0,T)=kN\ln\left(\frac{V}{V_0}\right)\,.
\end{equation}
For the relative entropies (\ref{eq:LQH-Step4}) of the radiation field, Boltzmann's
principle (\ref{eq:LQH-BoltzmanPrinzip}) now gives
\begin{equation}
\label{eq:LQH-Step6} W=\left(\frac{V}{V_0}\right)^{ E/h\nu}\,.
\end{equation}
From the striking similarity of (\ref{eq:LQH-BoltzmanAppl}) to (\ref{eq:LQH-Step6})
Einstein finally concludes:
\begin{quote}
\emph{``Monochromatic radiation of low density (within the range of Wien's radiation
formula) behaves thermodynamically as if it consisted of mutually independent energy quanta
of magnitude $h\nu$.''}
\end{quote}
So far no revolutionary statement has been made. The famous sentences just quoted express
the result of a statistical mechanical analysis.

\subsubsection*{Light quantum hypothesis}
Einstein's bold step consists in a statement about the quantum properties of the free
electromagnetic field, that was not accepted for a long time by anybody else. He formulates
his heuristic principle as follows:
\begin{quote}
\emph{``If, with regard to the dependence of its entropy on volume, a monochromatic
radiation (of sufficient low density) behaves like a discontinuous medium consisting of
energy quanta of magnitude $h\nu$, then it seems reasonable to investigate whether the laws
of generation and conversion of light are so constituted as if light consisted of such
energy quanta.''}
\end{quote}

In the final two sections, Einstein applies this hypothesis first to an explanation of
Stokes' rule for photoluminescence and then turns to the photoelectric effect. One should
be aware that in those days only some qualitative properties of this phenomenon were known.
Therefore, Einstein's well-known linear relation between the maximum kinetic energy of the
photoelectrons ($E_{\rm max}$) and the frequency of the incident radiation,
\begin{equation}
\label{eq:LQH-PhotoEff} E_{\rm max}=h\nu-P\,,
\end{equation}
was a true prediction. Here $P$ is the work-function of the metal emitting the electrons,
which depends on the material in question but not on the frequency of the incident light.

It should be stressed that Einstein's bold light quantum hypothesis was very far from
Planck's conception. Planck neither envisaged a quantization of the free radiation field,
nor did he, as it is often stated, quantize the energy of a material oszillator per se.
What he was actually doing in his decisive calculation of the entropy of a harmonic
oscillator was to assume that the \emph{total} energy of a large number of oscillators is
made up of \emph{finite} energy elements of equal magnitude $h\nu$. He did not propose that
the energies of single material oscillators are physically
quantized.\footnote{\label{foot:Planck} In 1911 Planck even formulated a `new radiation
hypothesis', in which quantization only applies to the process of light emission but not to
that of light absorption (Planck 1911). Planck's explicitly stated motivation for this was
to avoid an effective quantization of oscillator energies as a \emph{result} of
quantization of all interaction energies. It is amusing to note that this new hypothesis
led Planck to a modification of his radiation law, which consisted in the addition of the
temperature-independent term $h\nu/2$ to the energy of each oscillator, thus corresponding
to the oscillator's energy at zero temperature. This seems to be the first appearance of
what soon became known as `zero-point energy'.} Rather, the energy elements $h\nu$ were
introduced as a formal counting device that could at the end of the calculation not be set
to zero, for, otherwise, the entropy would diverge. It was Einstein in 1906 who interpreted
Planck's result as follows (CPAE, Vol.\,2, Doc.\,34):
\begin{quote}
\emph{``Hence, we must view the following proposition as the basis underlying Planck's
theory of radiation: The energy of an elementary resonator can only assume values that are
integral multiples of $h\nu$; by emission and absorption, the energy of a resonator changes
by jumps of integral multiples of $h\nu$.'' }
\end{quote}
\section{Energy and momentum fluctuations of the \\radiation field}
In his paper ``On the present status of the radiation problem'' of 1909 (CPAE, Vol.\,2,
Doc.\,56), Einstein returned to the considerations discussed above, but extended his
statistical analysis to the entire Planck distribution. First, he considers the energy
fluctuations, and re-derives the general fluctuation formula he had already found in the
third of his statistical mechanics articles. This implies for the variance of $E_V$ in
(\ref{eq:LQH-Energy}):
\begin{equation}
\label{eq:E-VarianceGeneral} \left\langle(E_V-\langle E_V\rangle)^2\right\rangle =
kT^2\frac{\partial\langle E_V\rangle}{\partial T}
=kT^2V\Delta\nu\frac{\partial\rho}{\partial T}\,.
\end{equation}
For the Planck distribution this gives
\begin{equation}
\label{eq:E-VariancePlanck} \left\langle(E_V-\langle E_V\rangle)^2\right\rangle
=\left(h\nu\rho+\frac{c^3}{8\pi\nu^2}\rho^2\right)V\Delta\nu\,.
\end{equation}
Einstein shows that the second term in this most remarkable formula, which dominates in the
Rayleigh-Jeans regime, can be understood with the help of the classical wave theory as due
to the interferences between the partial waves. The first term, dominating in the Wien
regime, is thus in obvious contradiction with classical electrodynamics. It can, however,
be interpreted by analogy to the fluctuations of the number of molecules in ideal gases,
and thus represents a particle aspect of the radiation in the quantum domain.

Einstein confirms this particle-wave duality, at this time a genuine theoretical conundrum,
by considering also the momentum fluctuations. For this he considers the Brownian motion of
a mirror which perfectly reflects radiation in a small frequency interval, but transmits
for all other frequencies. The final result he commented as follows:
\begin{quote}
\emph{``The close connection between this relation and the one derived in the last section
for the energy fluctuation is immediately obvious, and exactly analogous considerations can
be applied to it. Again, according to the current theory, the expression would be reduced
to the second term (fluctuations due to interference). If the first term alone were
present, the fluctuations of the radiation pressure could be completely explained by the
assumption that the radiation consists of independently moving, not too extended complexes
of energy $h\nu$.''}
\end{quote}

Einstein discussed these issues also in his famous Salzburg lecture (CPAE Vol.\,2,
Doc.\,60) at the 81st Meeting of German Scientists and Physicians in 1909. Pauli (1949)
once said that this report can be regarded as a turning point in the development of
theoretical physics. In this Einstein treated the theory of relativity and quantum theory
and pointed out important interconnections between his work on the quantum hypothesis, on
relativity, on Brownian motion, and statistical mechanics. Already in the introductory
section he says prophetically:
\begin{quote}
\emph{``It is therefore my opinion  that the next stage in the development of theoretical
physics will bring us a theory of light that can be understood as a kind of fusion of the
wave and emission theories of light''.}
\end{quote}
We now know that it took almost twenty years until this was achieved by Dirac in his
quantum theory of radiation.

\section{Reactions}

We already stressed that Einstein's bold light quantum hypothesis was very far from
Planck's conception. This becomes particularly evident from the following judgement of
Planck.

When Planck, Nernst, Rubens, and Warburg proposed Einstein in 1913 for membership in the
Prussian Academy their recommendation concludes as follows:
\begin{quote}
``\textit{In sum, one can say that there is hardly one among the great problems in which
modern physics is so rich to which Einstein has not made a remarkable contribution. That he
may sometimes have missed the target in his speculations, as, for example, in his
hypothesis of light-quanta, cannot really be held to much against him, for it is not
possible to introduce really new ideas even in the most exact sciences without sometimes
taking a risk.}''
\end{quote}

It took almost ten years until Einstein's application of the light quantum hypothesis to
the photoelectric effect was experimentally confirmed by Millikan, who then used it to give
a first precision measurement of $h$ (slope of the straight line given by
(\ref{eq:LQH-PhotoEff}) in the $\nu$-$E_{\rm max}$ plane) at the $0.5$ percent level
(Millikan 1916). Strange though understandable, not even he, who spent 10 years on the
brilliant experimental verification of its consequence (\ref{eq:LQH-PhotoEff}), could
believe in the fundamental correctness of Einstein's hypothesis. In his comprehensive paper
(Millikan 1916) on the determination of $h$, Millikan first commented on the light-quantum
hypothesis:
\begin{quote}
\emph{ ``This hypothesis may well be called reckless, first because an electromagnetic
disturbance which remains localized in space seems a violation of the very conception of an
electromagnetic disturbance, and second because it flies in the face of the thoroughly
established facts of interference.''}
\end{quote}
And after reporting on his successful experimental verification of Einstein's equation
(\ref{eq:LQH-PhotoEff}) and the associated determination of $h$, Millikan  concludes:
\begin{quote}
\emph{``Despite the apparently complete success of the Einstein equation, the physical
theory of which it was designed to be the symbolic expression is found so untenable that
Einstein himself, I believe, no longer holds to it.''}
\end{quote}

Most of the leading scientists (Sommerfeld, von Laue, Bohr, etc) strongly opposed
Einstein's idea of the light-quantum, or at least openly stated disbelief.
\section{Derivation of the Planck distribution}
A peak in Einstein's endeavor to extract as much as possible about the nature of radiation
from the Planck distribution is his paper ``On the Quantum Theory of Radiation'' of 1916
(CPAE, Vol.\,6, Doc.\,38). In the first part he gives a derivation of Planck's formula
which has become part of many textbooks on quantum theory. Einstein was very pleased by
this derivation, about which he wrote on August 11th 1916 to Besso: ``An amazingly simple
derivation of Planck's formula, I should like to say \emph{the} derivation''. For it he
introduced the hitherto unknown process of induced emission\footnote{Einstein's derivation
shows that without assuming a non-zero probability for induced emission one would
necessarily arrive at Wien's instead of Planck's radiation law.}, next to the familiar ones
of spontaneous emission and induced absorption. For each pair of energy levels he described
the statistical laws for these processes by three coefficients (the famous $A$- and
$B$-coefficients) and established two relations amongst these coefficients on the basis of
his earlier correspondence argument in the classical Rayleigh-Jeans limit and Wien's
displacement law. In addition, the latter also implies that the energy difference
$\varepsilon_n-\varepsilon_m$ between two internal energy states of the atoms in
equilibrium with thermal radiation has to satisfy Bohr's frequency condition:
$\varepsilon_n-\varepsilon_m=h\nu_{nm}$. In Dirac's 1927 radiation theory these results
follow ---without any correspondence arguments---from first principles.

In the second part of his fundamental paper, Einstein discusses the exchange of momentum
between the atoms and the radiation by making use of the theory of Brownian motion. Using a
truly beautiful argument he shows that in every elementary process of radiation, and in
particular in spontaneous emission, an amount $h\nu/c$ of momentum is emitted in a random
direction and that the atomic system suffers a corresponding recoil in the opposite
direction. This recoil was first experimentally confirmed in 1933 by showing that a long
and narrow beam of excited sodium atoms widens up after spontaneous emissions have taken
place (R.\,Frisch 1933). Einstein's paper ends with the following remarkable statement
concerning the role of ``chance'' in his description of the radiation processes by
statistical laws, to which Pauli (1948) drew particular attention:
\begin{quote}
\emph{``The weakness of the theory lies, on the one hand, in the fact that it does not
bring us any closer to a merger with the undulatory theory, and, on the other hand, in the
fact that it leaves the time and direction of elementary processes to `chance'; in spite of
this I harbor full confidence in the trustworthiness of the path entered upon.''}
\end{quote}

\section{Bose-Einstein statistics for degenerate material gases}
The last major contributions of Einstein to quantum theory were stimulated by de\,Broglie's
suggestion that material particles have also a wave aspect, and Bose's derivation of
Planck's formula that made only use of the corpuscular picture of light, though based upon
statistical rules using their indistinguishability. Einstein applied Bose's statistics for
photons to degenerate gases of identical massive particles. With this `Bose-Einstein
statistics', he obtained a new law, to become known as the Bose-Einstein distribution. As
for radiation, Einstein considered again fluctuations of these gases and found both,
particle-like and wave-like aspects. But this time the wave property was the novel feature,
that was recognized by Einstein to be necessary.

In the course of this work on quantum gases, Einstein discovered the condensation of such
gases at low temperatures. (Although Bose made no contributions to this, one nowadays
speaks of Bose-Einstein condensation.) Needless to say that this subject has become
enormously topical in recent years.

Schr\"{o}dinger acknowledged in his papers on wave mechanics the influence of Einstein's
gas theory, which from todays perspective appear to be his last great contribution to
physics. In the article in which Schr\"{o}dinger (1926) establishes the connection of
matrix and wave mechanics, he remarks in a footnote: ``My theory was inspired by
L.\,de\,Broglie and by brief but infinitely far-seeing remarks of A.\,Einstein
(\textit{Berl}. \textit{Ber}. 1925, p. 9ff)''.

It is well-known that Einstein considered the `new' quantum mechanics less than
satisfactory until the end of his life. In his autobiographical notes he says, for example,
\begin{quote}
\emph{I believe, however, that this theory offers no useful point of departure for future
developments. This is the point at which my expectation departs widely from that of
contemporary physicists.''}
\end{quote}

\section{Einstein and the interpretation of \\quantum mechanics} \label{sec:QMinterp}
The new generation of young physicists that participated in the tumultuous three-year
period from January 1925 to January 1928 deplored Einstein's negative judgement of quantum
mechanics. In his previously cited article on Einstein's contributions to quantum
mechanics, Pauli (1949) expressed this with the following words:
\begin{quote}
\emph{``The writer belongs to those physicists who believe that the new epistemological
situation underlying quantum mechanics is satisfactory, both from the standpoint of physics
and from the broader knowledge in general. He regrets that Einstein seems to have a
different opinion on this situation (...).''}
\end{quote}
When the Einstein-Podolski-Rosen (EPR) paper (Einstein \emph{et\,al}.~1935) appeared,
Pauli's immediate reaction (see Pauli 1985-99, Vol.\,2) in a letter to  Heisenberg of June
15th was quite furious:
\begin{quote}
\textit{``Einstein once again has expressed himself publicly on quantum mechanics, namely
in the issue of Physical Review of May 15th (in cooperation with Podolsky and Rosen -- not
a good company, by the way). As is well known, this is a catastrophe each time when it
happens.''}
\end{quote}
From a greater distance in time this judgement seems exaggerated, but it shows the attitude
of the `younger generation' towards Einstein's concerns. In fact, Pauli understood (though
not approved) Einstein's point much better than many others, as his intervention in the
Born-Einstein debate on Quantum Mechanics shows (Born 2005, letter by Pauli to Born of
March 31st 1954). Whatever one's attitude on this issue is, it is certainly true that the
EPR argumentation has engendered an uninterrupted discussion up to this day. The most
influential of John Bell's papers on the foundations of quantum mechanics has the title
``On the Einstein-Podolsky-Rosen paradox'' (Bell 1964). In this publication Bell presents
what has come to be called ``Bell's Theorem'', which (roughly) asserts that \emph{no
hidden-variable theory that satisfies a certain locality condition can produce all
predictions of quantum mechanics}. This signals the importance of EPR's paper in focusing
on a pair of well-separated particles that have been properly prepared to ensure strict
correlations between certain of their observable quantities. Bell's analysis and later
refinements (1987) showed clearly that the behavior of entangled states is only explainable
in the language of quantum mechanics.

This point has also been the subject of the very interesting, but much less known work of
S.\,Kochen and E.P.\,Specker (1967), with the title ``The Problem of Hidden Variables in
Quantum Mechanics''. Loosely speaking, Kochen and Specker show that quantum mechanics
\emph{cannot} be embedded into a classical stochastic theory, provided two very desirable
conditions are assumed to be satisfied. The first condition (KS1) is that the quantum
mechanical distributions are reproduced by the embedding of the quantum description into a
classical stochastic theory. (The precise definition of this concept is given in the cited
paper.) The authors first show that hidden variables in this sense can always be introduced
if there are no other requirements. (This fact is not difficult to prove.) The second
condition (KS2) imposed by Kochen and Specker states that a function $u(A)$ of a quantum
mechanical observable $A$ (self-adjoint operator) has to be represented in the classical
description by the very same function $u$ of the image $f_A$ of $A$, where $f$ is the
embedding that maps the operator $A$ to the classical observable $f_A$ on `phase space'.
Formally, (KS2) states that for all $A$
\begin{equation}
\label{eq:KS2} f_{u(A)}=u\left(f_A\right).
\end{equation}

The main result of Kochen and Specker states that if the dimension of the Hilbert space of
quantum mechanical states is larger than 2, an embedding satisfying (KS1) and (KS2) is `in
general' \emph{not possible}.

There are many highly relevant examples---even of low dimensions with only a finite number
of states and observables---where this impossibility holds.

The original proof of Kochen and Specker is very ingenious, but quite difficult. In the
meantime several authors have given much simpler proofs; e.g. Straumann (2002).

We find the result of Kochen and Specker entirely satisfactory in the sense that it clearly
demonstrates that there is no way back to classical reality. Einstein's view that quantum
mechanics is a kind of glorified statistical mechanics, that ignores some hidden
microscopic degrees of freedom, can thus not be maintained without giving up locality or
(KS2). It would be interesting to know his reaction to these developments that have been
triggered by the EPR paper.

Entanglement is not limited to questions of principle. It has already been employed in
quantum communication systems, and entanglement underlies all proposals of quantum
computation.

\subsection*{Acknowledgements}
I sincerely thank Domenico Giulini for a fruitful collaboration on an extensive paper
devoted to ``Einstein's Impact on the Physics of the Twentieth Century'', to appear in
``Studies in History and Philosophy of Modern Physics''.

\end{document}